\documentclass[onecolumn,showpacs,amsmath,amssymb,prd,nofootinbib]{revtex4-2}
\usepackage{graphicx}
\usepackage{epsfig}
\usepackage{enumerate}
\def\sss{\scriptscriptstyle}
\def\^#1{^{\sss #1}}
\def\_#1{_{\sss #1}}
\def\av#1{\langle #1\rangle}
\def\beq{\begin{equation}}
\def\eeqno#1{\label{#1}\end{equation}}

\def\az{a_{\scriptscriptstyle 0}}
\def\A{\mathcal{A}}

\def\azg{\A_0}

\def\rar{\rightarrow}
\def\s{\sigma}
\def\r{\rho}
\def\l{\lambda}

\def\drt{d^3r}
\def\f{\phi}
\def\grad{\vec\nabla}
\def\div{\vec \nabla\cdot}
\def\gf{\grad\phi}

\def\vr{ \textbf{r}}

\def\vv{\textbf{v}}
\def\va{\textbf{a}}
\def\vg{\textbf{g}}

\def\vk{\textbf{k}}
\def\vF{\textbf{F}}
\def\o{\omega}

\def\L{\mathcal{L}}
\def\fN{\f\_N}

\def\m{\mu}
\def\a{\alpha}
\def\b{\beta}
\def\c{\gamma}
\def\d{\delta}

\def\k{\kappa}

\def\D{\Delta}
\def\fpg{4\pi G}

\def\gN{g_{\scriptscriptstyle N}}
\def\vgN{\vg_{\scriptscriptstyle N}}
\def\vinf{V_\infty}

\def\S{\Sigma}

\def\rM{r\_M}
\def\SB{\Sigma\_B}
\def\SM{\Sigma\_M}
\def\SD{\Sigma\_D}
\def\SDz{\S\^0\_D}
\def\SBz{\S\^0\_B}
\def\Lap#1{\D^{[#1]}}

\begin{document}
\title{The deep-MOND limit -- a study in Primary vs secondary predictions\footnote{Bases on a talk presented at the MOND workshop, Leiden, September 2025.}}

\author{Mordehai Milgrom}
\affiliation{Department of Particle Physics and Astrophysics, Weizmann Institute}

\begin{abstract}
In default of a fundamental MOND theory -- a FUNDAMOND -- I advocate that, alongside searching for one, we should try to identify predictions that follow from wide classes of MOND theories, if not necessarily from all. In particular, predictions that follow from only the basic tenets of MOND -- ``primary predictions'' -- are shared by all MOND theories, and are especially valuable. Such predictions permit us to test the MOND paradigm itself, or at least large parts of it, without yet having a FUNDAMOND. Concentrating on the deep-MOND limit, I discuss examples of either type of predictions. For some examples of primary predictions, I demonstrate how they follow from the basic tenets (which I first formulate). I emphasize that even predictions that pertain to the deep-MOND limit - namely, those that concern gravitating systems that have low accelerations everywhere -- require the full set of MOND tenets, including the existence of a Newtonian limit close to the deep-MOND regime. This is because Newtonian dynamics is a unique theory that all MOND theories must tend to in the limit of high accelerations, and it strongly constrains aspects of the deep-MOND regime, if the transition between the limits is fast enough, which is one of the MOND tenets.
\end{abstract}
\maketitle

\section{Introduction}
This workshop's aim is mainly to propose and discuss future paths -- both theoretical and observational -- for advancing the MOND research program\footnote{For reviews of MOND with various emphases, see Refs. \cite{fm12,milgrom14,milgrom20a,mcgaugh20,merritt20,bz22,fd25}.}. To do this efficiently, we have to look into the past, and take stock of what has been achieved so far. A fact that strikes one is that we do not yet have a fully satisfactory MOND theory.
\par
It is obvious then, that important future endeavours will involve improving on existing theories and devising new, more appealing ones.
\par
However, a useful path that should be pursued in parallel with theory development, is deducing theory-agnostic predictions from only some  basic MOND tenets (see Ref. \cite{milgrom14a,milgrom24} for a detailed account of this approach). We refer to these as ``primary'' or ``first-tier'' predictions. It is important to identify such predictions, and contrast them with ``secondary'' phenomena, for which MOND predictions do depend on the specific MOND theory. Such predictions can, sometimes, differ greatly among MOND theories that, otherwise, make the same primary predictions (see Ref. \cite{milgrom23} for several such examples).
\par
It is important to make this distinction for understanding what exactly it is that we are testing with the various observations and measurements. Is it the specific theory whose prediction we employ, or is it the MOND paradigm itself? Having such distinction in mind will also direct us in picking the tests we want to conduct, and is also good to keep in mind when building theories.
\par
In fact, an even finer classification of MOND predictions is called for, inasmuch as there are prediction that, while not primary, follow from a large class of theories, but still may differ from those of other classes.
\par
Here, I limelight some of these points with specific examples of MOND predictions that apply to nonrelativistic (NR), self-gravitating systems that are governed by the deep-MOND limit (DML). Namely, systems -- such as low-surface-density disc galaxies, dwarf spheroidal satellite of larger galaxies, many galaxy groups, etc. -- where all the relevant accelerations are much smaller than $\az$, the MOND acceleration constant.
\par
In Sec. \ref{axioms}, I lay out the standard set of MOND's basic tenets, with emphasis on the definition of the dynamics in the DML. Section \ref{MGMI} explains what I mean by classifying NR theories as ``modified gravity'' (MG), ``modified inertia'' (MI), and more generally, theories that modify both gravity and inertia. In Sec. \ref{SIsec}, I employ scale invariance -- the essential symmetry underlying the DML -- to derive the scaling relations that follow from it. In Sec. \ref{toy}, I describe a family of DML models that serve as a heuristic tool for demonstrating some of the general points I am trying to bring home. Examples of observational consequences of these relations are then derived in Sec. \ref{consequences}, and comparisons are made between the predictions of various MOND theories.

\section{MOND -- basic tenets \label{axioms}}
Deep-MOND-limit dynamics is sometimes described as encapsulated by the relations $g=(\az\gN)^{1/2}$, between the Newtonian acceleration, $\gN$, derived from the baryonic mass distribution, and the acceleration, $g$, predicted by MOND for that distribution. If generally valid, this would have been a very powerful relation, since it would give us the MOND acceleration straightforwardly. However, such a relation cannot be taken to describe the DML. First, it is not well defined, because it does not give the direction of the MOND acceleration. We could try to take  a vectorial relation $\vg=(\az/\gN)^{1/2}\vgN$. But this would be inadequate, for example, because in existing MOND theories the MOND acceleration does not even vanish at the same points as $\vgN$ does.
Also, as pointed out at the outset, such a relation might at best be applied to the dynamics of test particles in a given gravitational field, because it breaks momentum conservation, for example, and leads to self-acceleration of isolated systems, if applied generally to many-body systems.
\par
We thus need a proper definition of what defines DML dynamics. This is done within the general formulation of the basic MOND tenets.
\par
I have argued on many occasions (e.g., in Refs. \cite{milgrom14,milgrom20a,milgrom20b}) that understanding and formulating MOND in a relativistic framework may require completely new concepts that are, at present, beyond our ken. For example, the ``coincidence'' of the MOND constant, $\az$, with cosmologically significant accelerations, makes cosmology the only strong-gravity relativistic system in the MOND regime. This may point to MOND having to be understood together with cosmology as inseparable aspects of the same FUNDAMOND, and not as some relativistic meta-theory for which cosmology is only a special solution.
\par
For this reason, I have avoided formulating general basic tenets for relativistic MOND theories, except that they have to reduce to a theory that reproduces MOND phenomenology for NR systems.
\par
The basic tenets as I see then now, formulated for the NR regime are as follows.
\par
(i) MOND is a theory of dynamics (gravity plus inertia) involving a new constant, $\az$.
\par
(ii) For systems where all attributes with the dimensions of acceleration are much larger than $\az$, Newtonian dynamics is restored as a limit. Formally, this is expressed as the fact that any MOND relation in which $\az$ appears tends to the corresponding Newtonian relation in the limit $\az\rar 0$.
\par
(iii) In the opposite, DML, where all relevant accelerations in the system are much smaller than $\az$, dynamics become spacetime scale invariant (SI) \cite{milgrom09}. This limit is achieved formally by taking in a MOND relation, $\az\rar \infty$, and $G\rar 0$, while $\azg\equiv G\az$ is kept fixed. In this limit the MOND relation that is gotten should be invariant to scaling of all lengths, and all times, by a (positive) factor $\l$:
\beq (t,\vr)\rar\l(t,\vr). \eeqno{scooka}
\par
(iv) As a minimalist requirement -- and until forced to proceed otherwise -- we should require that no other new dimensionful constant should appear in the theory, besides $\az$, neither should the theory involve dimensionless constants that are not of order unity. This means, e.g., that $\az$ determines not only the value of the Newtonian-to-MOND-transition acceleration, but also the width of the transition.
\par
There are additional -- non-MOND-specific -- requirements from a MOND theory. For example:
\par
(v) The standard conserved quantities and their conservation laws (energy, momentum, and angular momentum). In action-based theories these follow from the appropriate symmetries of the action.
\par
(vi) We should require some form of boost invariance -- like Galilei invariance in Newtonian dynamics, and Lorentz invariance in standard relativity -- lest, for example, the internal dynamics of two identical galaxies would differ if they move with respect to each other. Reference \cite{milgrom94} (Sec. III there) brings additional reasons to require some form of boost invariance.
\par
(vii) It is sensible to require universality of free fall in a gravitational field even in the DML. This principle has not been tested in the DML to the same high accuracy that it is known to hold in high-acceleration systems, such as the solar system. However, observations tell us that, even in the DML, different types of test bodies (e.g., stars of all types, gas clouds, etc.) move, as best as we can tell, in the same way in the fields of galaxies. Note, importantly, that this does not mean that all bodies are subject to the same acceleration at a given position in a gravitational field. It only means that the physical trajectories in the field are independent of the type of test body.
\par
(viii) Specifically, in the context of MOND, we require that composite bodies move on the same trajectories whatever the internal accelerations of their constituents are (high or low in the MOND sense). This is known as the requirement of a correct center-of-mass motion of composite bodies.
\par
(ix) We also want the dynamics of test bodies asymptotically far from a bound and isolated massive body to depend only on the total mass of the body, and not on details of its structure. This may not be an independent requirement, which needs to be further checked.
\par
One may, perhaps come up with additional requirements according to ones conception of a physical theory.
\section{Between modified gravity and modified inertia \label{MGMI}}
A possible classification of MOND theories is according to whether they modify only Newtonian and general-relativistic gravity, or whether they modify also the matter sector -- in the extreme case whether they modify only inertia. In general -- especially in the case of relativistic theories, the classification is not clear-cut, and may not even be so useful \cite{milgrom23}. However, in the NR domain -- which is relevant to galactic dynamics -- the classification is rather clear-cut and useful.
\par
Since the difference between MOND and Newtonian dynamics is brought to bear mainly in the DML, consider first the schematic (e.g., suppressing the vectorial nature of the relations), mutual scaling between acceleration, $a$, active gravitational mass\footnote{I assume that the theories we want to construct obey the universality of free fall; so the inertial mass and the passive gravitational mass do not appear.}, $M$, and distance, $r$.
The Newtonian relation, $a=MG/r^2$, is clearly not SI, and there are many ways that involve $\az$ to modify it so as to obtain a SI relation. For example, we can schematically modify it to
\beq a^2/\az=MG/r^2,  \eeqno{inertia}
which could be viewed as pure MI if the right-hand side is still taken to express the force per unit mass, and the left-hand side is taken as the inertia term. Or, at the other extreme, we could schematically modify it to the equivalent expression
\beq a=(M\azg)^{1/2}/r, \eeqno{gravity}
which would constitute a modified-gravity relation, if the same roles of the two sides of the relation are preserved. But, there is no reason why the only attribute of the particle trajectory that gravity dictates is the acceleration. So, more general SI relations between mass, $M$, trajectory, $\vr(t)$, and time, $t$, can be constructed\footnote{We would then require, in addition to rotation and space translation invariance, also invariance to time translation.}.
\par
For example, a schematic relation of the form
\beq \left(\frac{d^\a r}{dt^\a}\right)^\b=\frac{(M\azg)^{\a\b/4}}{r^{\b(\a-1)}}  \eeqno{mixed}
is SI  \cite{milgrom25}.
As before, with the interpretation of the left-hand side as inertia, and of the right-hand side as gravity, this would be understood as modification of both Newtonian inertia and gravity. Note that for noninteger $\a$, the inertia term is time nonlocal, since fractional time derivatives are defined in terms of the Fourier components of the trajectory; see details in Ref. \cite{milgrom25}.
\par
More formally, for a NR, self-gravitating system, of masses $m\_p$, on trajectories $\vr\_p(t)$, start with the Newtonian action
$I=\int L~dt$, where the Lagrangian, $L$, is
\beq L= \sum_p m\_p[\frac{1}{2} \vv\_p^2-\f(\vr\_p)]-\frac{1}{8\pi G}\int d^3x~(\gf)^2.   \eeqno{newtlag}

The equations of motion of the masses and the field equation for the gravitational potential are
\beq \ddot\vr\_p=-\gf(\vr\_p),~~~~~~~~~~~~\Delta\f=\fpg\rho,~~~~~{\rm where}~~~~~~~\r(\vr,t)=\sum_p m\_p\d[\vr-\vr\_p(t)].   \eeqno{equat}

In MG theories, we modify the last (gravitational) term to get \cite{milgrom14b})
 \beq L=\sum_p m\_p[\frac{1}{2} \vv\_p^2-\f(\vr\_p)]- L_f(\f,\psi,G,\az), \eeqno{kalur}
 where, the first of Eqs. (\ref{equat}) still holds, so $\f$ is the gravitational potential, but the gravitational Lagrangian can be a general functional of the field ${\f(\vr)}$ and of other gravitational degrees of freedom -- possibly tensorial -- collectively designated $\psi$.
\par
Restricting ourselves to local gravitational Lagrangians, we take
 \beq  L_f=\int  \L_f(\f,\psi,G,\az)~\drt.  \eeqno{bufa}
\par
The MG, MOND theories that have been propounded to date (e.g., AQUAL \cite{bm84}, QUMOND \cite{milgrom10}, and TRIMOND \cite{milgrom23b}), are of this class.
\par
It was shown in Ref. \cite{milgrom14b} that this class of theories predict for an isolated, self-gravitating, DML system of ``point'' bodies of masses $m\_p$, at positions $\vr\_p$, a general relation of the form
\beq \sum_p \vr_p\cdot\vF_p=-\frac{2}{3}\azg^{1/2}[M^{3/2}-\sum_p m_p^{3/2}], \eeqno{i}
where $\vF_p$ are the gravitational forces acting on the bodies, and $M=\sum_p m_p$. The accelerations the bodies are subject to are $\va\_p=\vF_p/m\_p$.
\par
From relation (\ref{i}) follow several important predictions. For example, the DML force between two arbitrary ``point'' masses, a distance $\ell$ apart, is
 \beq F(m_1,m_2,\ell)=\frac{2}{3}\frac{\azg^{1/2}}{\ell}[(m_1+m_2)^{3/2}-m_1^{3/2}-m_2^{3/2}]. \eeqno{twobody}
\par
Another prediction is a DML virial relation that applies to isolated, DML systems whose structure is stationary -- i.e., whose structure changes only little on dynamical time scales:
\beq  \s^2=\frac{2}{3}(MG\az)^{1/2}[1-\sum_p (m_p/M)^{3/2}],   \eeqno{jatare}
where $\s^2=M^{-1}\sum_p m_p\vv_p^2$ is the mass-weighted mean squared velocity in the system\footnote{If the system is not stationary, but remains bounded, then $\s^2$ should be understood as the long-time average.}.
\par
These are examples of important predictions that follow in a large class of MG theories.
\par
At the other extreme, MI, in the restricted sense that I adopt here, modifies only the particle kinetic term in the action, $m\_p\int (\vv^2/2)dt$, replacing it with some functional of the particle trajectory of the form $m\_p F[\vr\_p(t),\az]$. The proportionality to $m\_p$ is retained so as not to violate the universality of free fall in a gravitational field. More details, with examples, can be found in Refs. \cite{milgrom94,milgrom23}.
\par
Examples of mixed MG-MI models are described in Ref. \cite{milgrom25}, and below, in Sec. \ref{toy}.

\section{General scaling relations of the DML  \label{SIsec}}
Quantities such as lengths, time, velocities, kinematic accelerations, $d^2\vr/dt^2$, fractional time derivatives of the trajectories, etc. transform under the spacetime scaling (\ref{scooka}) with a power of $\l$ -- the ``scaling dimension'' -- that matches their length and time dimensions. This holds, more generally, if a degree of freedom, $\psi$, has dimensions $[m]^a[\ell]^b[t]^c$, and it has scaling dimension $b+c$. This is not always the case with the standard definitions of the degrees of freedom. For example, the Newtonian potential, $\fN$, has dimensions of velocity, $b=-c=2$, but scaling dimension $-1$ (as it scales as $MG/r$), not $0$.
But, given a theory that obeys the DML tenets, we can always, as an auxiliary step, give this property to a degree of freedom by multiplying it by a power of $\az$.
For example, $\fN$ appears in QUMOND; so we write the theory with the standardized potential $\psi\equiv \az\fN$, whose dimensions do match its scaling dimension. A MOND potential does have this property.
\par
Writing the DML theory in terms of such standardized degrees of freedom, SI implies that the constants $\az$ and $G$ can only appear in the theory as their product $\azg= G\az$. This has to do with the fact that under a {\it change of the units}\footnote{Not to be confused with spacetime scaling, under which constants do not change.} of length and time by same factor, the values of $\az$ and $G$ change, while that of $\azg$ does not, as its dimensions are
\beq   [\azg]=[m]^{-1}[\ell]^4[t]^{-4}.   \eeqno{azgdim}
(See, e.g.,  Refs. \cite{milgrom09,milgrom14a} for discussions of this result from different angles.)
The opposite is also true: If the values of all the dimensionful constants that appear in a theory do not change under the simultaneous rescaling of the time and length units by the same factor, then the theory is SI.
\par
However, the dimensions of $\azg$ being what they are, its value does not change under a larger, two-parameter family of changes of units that correspond to a change of values\footnote{The units of mass is scaled by $\k^{-1}$, hence the values of masses are multiplied by $\k$, etc.}
\beq  m\rar \k m,~~\ell\rar\l\ell,~~~t\rar \c t\equiv \l\k^{-1/4}t.   \eeqno{units}
And, since $\azg$ is {\it the only dimensionful constant} that can appear in a (standardized) DML theory, the theory must be invariant to the larger, two-parameter family of scalings\footnote{If another constant appears -- for example a speed -- whose value is also invariant to the same scaling of the units of time and length, but with a different mass dimensions from $\azg$, we would still have SI, but not the two-parameter scaling invariance.}, which for an isolated system of point masses\footnote{If the bodies are not pointlike, their internal structure and dynamics have also to be in the DML, and need to be scaled.}, $m\_p$, whose trajectories are $\vr\_p(t)$, is
$$ m_p\rar \k m_p,~~~~\vr_p(t)\rar \l\vr_p(t/\c),~~~~\vv_p(t)\rar \k^{1/4}\vv_p(t/\c)$$
\beq \va_p(t)\rar \k^{1/2}\l^{-1}\va_p(t/\c),~~~~\vg^N_p(t)\rar \k\l^{-2}\vg^N_p(t/\c).  \eeqno{scasca}
Dynamical surface densities, which are deduced from observed accelerations, scale like the kinematic accelerations, $\va_p(t)$, and baryonic surface densities, which are deduced from observed baryonic masses, scale like the Newtonian, gravitational accelerations, $\vg^N_p(t)$. Scale invariance is described by the special case $\k=1$.
\par
In the continuum description, where the system is described by the density and velocity fields $\r(\vr,t)$ and $\vv(\vr,t)$, these scalings read
$$ \r(\vr,t)\rar  \k\l^{-3}\r(\vr/\l,t/\c),~~~\vv(\vr,t)\rar  \k^{1/4}\vv(\vr/\l,t/\c),$$
\beq \va(\vr,t)\rar  \k^{1/2}\l^{-1}\va(\vr/\l,t/\c),
~~~~\vg^N(\vr,t)\rar \k\l^{-2}\vg^N(\vr/\l,t/\c).  \eeqno{scabla}
Dynamical masses scale as $\k^{1/2}\l$.
\par
These predicted, DML scaling relations follow because applying the unit changes to a DML equation has the same effect as applying these scalings to only the degrees of freedom, since the value of $\azg$ is not affected. Note that the scalings relations do not require the standardization of the degrees of freedom, which does not change their scaling properties. I used it only to facilitate the derivation.
\par
In many instances, we apply these relations to stationary systems, such as galaxies, whose structure evolves on time scales much longer then their dynamical times. In these cases, we can ignore the time dependence, and hence the scaling of time in the arguments.
\section{Family of DML models  \label{toy}}
Before continuing with the discussion of observational consequences of the above scaling laws, I present briefly a one-parameter family of DML models, governing systems of pointlike bodies, that I shall use below for demonstrating various points. These are described in detail in Ref. \cite{milgrom25}. The Lagrangians of these models are
\beq  L= \frac{\azg}{2\b q(\b)}\sum_pm\_p [|\vr\_p^{(2/\b)}|^{2\b}-\psi(\vr\_p)]  -\int d^3r (\Lap{(2\b-1)/4}\psi)^2,  \eeqno{totaya}
for $1\le\b\le 2$. The numerical factor $q(\b)$ has an analytic expression (given in Ref. \cite{milgrom25}) that fixes the normalization of $\azg$ so that the mass-asymptotic-speed relation (MASR) is $\vinf^4=M\azg$. (Since only $\azg$ appears as a dimensionful constant, these models are DML theories, and the predicted MASR is ensured from the outset.)
\par
Briefly, the main properties of these toy models are as follows:
The gravitational part involves fractional Laplacians, and the inertial term involves fractional time derivatives (for $\b\not = 1,2$). Such fractional-derivative operations, are defined in Ref. \cite{milgrom25} in terms of the Fourier transforms of the differentiated quantities:
If
 \beq \psi(\vr)=\frac{1}{(2\pi)^3}\int d^3k~ \hat\psi(\vk){\rm e}^{i\vk\cdot\vr} ,  \eeqno{foutranas}
then
\beq \Lap{\a}\psi(\vr)\equiv-\frac{1}{(2\pi)^3}\int d^3k~|\vk|^{2\a}\hat\psi(\vk){\rm e}^{i\vk\cdot\vr}d^3k.  \eeqno{foutra}
And, if
\beq \vr(t)=\frac{1}{2\pi}\int_{-\infty}^{\infty}d\o~\hat\vr(\o)e^{i\o t}, \eeqno{gumcaap}
then
\beq \vr^{(\eta)}\equiv  \frac{1}{2\pi}\int_{-\infty}^{\infty}d\o~|\o|^\eta e^{i[\o t+\eta\frac{\pi}{2}s(\o)]}\hat\vr(\o),  \eeqno{fourapi}
where $s(\o)$ is the sign of $\o$.
\par
These models thus modify both gravity ($\psi$ is not $\fN$) and inertia. The inertia is nonlocal (for $\b\not = 1,2$), and nonlinear (for $\b\not = 1$).
The gravitational part is linear in the mass distribution $\r$, and, in fact, the potential has an analytic expression. It solves the (nonlocal for $\b\not = 3/2$) differential equation
\beq \Delta^{[\b-1/2]}\psi=\frac{\azg}{2q(\b)}\rho,  \eeqno{darati}
with the solution that vanishes at infinity (for $\b<2$) being
\beq  \psi(\vr)=-\frac{\azg}{4-2\b}\int d^3r'\frac{\rho(\vr')}{|\vr-\vr'|^{4-2\b}}.  \eeqno{nauo}
(The case $\b=2$ is also well defined -- see below.)
Except for the $\b=3/2$ case, the standard shell theorems do not hold.
\par
The models are Galilei invariant for $\b<2$ [$\vr^{(\eta)}$ for $\eta>1$ is invariant under $\vr(t)\rar \vr(t)+\vv\_0 t$].
\par
{\it Importantly, I could not sensibly extend these models to the Newtonian regime, so they remain pure DML models without a full MOND extent.}
\par
In addition, I have not examined the issue of the center-of-mass motion of composite bodies [requirement (viii) in Sec. \ref{axioms}] for $\b\not =1$.
\par
The particle equation of motion is of the form
\beq  \vec\A[\{\vr_p(t)\}]=\grad\psi(\vr_p),  \eeqno{partieq}
where $\vec\A[\{\vr_p(t)\}]$ is a nonlocal functional of the trajectory, with no dimensionful constants appearing.
\par
The expression for the rotational speed on a circular orbit in the midplane of an axisymmetric, plane symmetric mass distribution (an idealized rotation curve) is
\beq  V^4(r)=r^{5-2\b}|\partial\psi/\partial r|.   \eeqno{sarmil}
\subsection{Special cases}
The equation of motion for the model with $\b=2$ is
\beq  \frac{d(v^2\vv)}{dt}=v^2\va+2(\vv\cdot\va)\vv=\azg\int d^3r'\frac{\rho(\vr')(\vr-\vr')}{|\vr-\vr'|^2}. \eeqno{katew}
It is not Galilei invariant.
\par
In the model with $\b=3/2$, we have $\psi=\az\fN$, the standardized form of the Newtonian potential. It thus satisfies the usual shell theorems. It is a pure MI theory, and thus, from the general theorem \cite{milgrom94} [seen also from Eqs. (\ref{nauo}) and (\ref{sarmil})], the model predicts
\beq V^2(r)/r=a(r)=[\az\gN(r)]^{1/2},  \eeqno{munater}
for the ideal rotation curve.
\par
The model with $\b=1$ is a linear DML theory, with the particle equation of motion
\beq  \frac{d^4\vr}{dt^4}=\azg\int d^3r'\frac{\rho(\vr')(\vr-\vr')}{|\vr-\vr'|^4}.  \eeqno{lilalil}
Being linear, the model satisfies the requirement from the center-of-mass motion of composite bodies, and {\it it does not exhibit an external-field effect}.

\section{Some consequences  \label{consequences}}
Asymptotic flatness of rotational speeds, and the MASR for isolated systems, $\vinf^4=M\azg$, which underlies the observed ``baryonic Tully-Fisher relation'' (see, e.g., Refs. \cite{lelli19,mistele24}), is a universal prediction of MOND, as it follows from only SI [and the requirement that asymptotic dynamics depend only on the systems total mass - (ix) in Sec.\ref{axioms}].
\par
There are, however additional primary DML predictions whose derivation is more subtle, and which require more elaboration.
\par
Consider a stationary, isolated mass distribution in the DML\footnote{If the system is not isolated, but falls in some external, DML field, this field, or its sources, has to be scaled appropriately in all the applications below.}, to take a concrete example, an isolated, axisymmetric, disc galaxy with an exponential surface-density distribution, of central surface density $\S_0$, of scale length $h$, of scale height $z$, and some velocity distribution $\vv(\vr)$,  consistent with stationarity. Pick some attributes of the galaxy as follows: any acceleration, $a$ (e.g., the radial acceleration at three scale lengths, and half the scale height); some baryonic (Newtonian) acceleration, $\gN$, (e.g., the vertical component at two scale lengths, and one scale height); some velocity, $V$, (e.g., the mass-weighted RMS velocity over the galaxy); a mass, $m$ (e.g., the total mass within half the scale height); some dynamical surface density\footnote{``Dynamical density, surface density, or mass'', are the corresponding quantities that would be required in Newtonian dynamics to explain the observed kinematics. In the dark-matter paradigm they stand for the baryon-plus-dark-matter quantities. In MOND they are only auxiliary quantities, do not represent real quantities, can be negative, and may even not be well defined -- e.g., in MI, MOND theories \cite{milgrom23}.}, $\SD$; and, some baryonic surface density, $\SB$ (e.g., the average surface density within the scale length).
\par
Define, as three examples, the quantities
\beq  \eta_V\equiv V^4/m\azg,~~~~\eta\_\S\equiv \SD/(\SB\SM)^{1/2},~~~~\eta_a\equiv a/(\az\gN)^{1/2}. \eeqno{conseq}
They are dimensionless, and are invariant under the scalings (\ref{scasca}) or (\ref{scabla}). ($\SM\equiv \az/2\pi G$ is the MOND critical surface density.)\footnote{$\SD$ is determined from kinematic accelerations, $a$, through expressions that scale as $a/G$. Thus schematically, $\eta\_\S\propto a/(\azg\SB)^{1/2}$.}
Thus, MOND predicts that the values of these quantities are universal within the two-parameter family of galaxies generated from the starting one by applying the scaling. In the above example, they are universal among all DML, exponential discs, with the same $z/h$ ratio.
This intra-family universality in itself, is an unavoidable, primary prediction. However, the numerical values these constants take may differ greatly among MOND theories, depending on the exact choice of the attribute that go into their construction. Another drawback of such general definitions is that they apply to only a restricted family self-similar objects; so they do not afford general application as the $M-\vinf$ relation does.
\par
For a given theory, there may be a definition of some of the $\eta$ parameters that are predicted to have a universal value, at least within a large class of galaxy types. Some examples are as follows:

(a) Clearly, If $V$ is taken as $\vinf$ and $m$ is the total mass, then, all MOND theories predict $\eta\_V=1$.

(b) For all ``pure MI'' theories, if $a$ is the radial, MOND acceleration, at $r$, in the midplane of any axisymmetric, plane-reflection symmetric system, and $\gN$ is the baryonic, Newtonian acceleration at $r$, then $\eta_a=1$ is predicted \cite{milgrom94}.

(c) In the large class of MG theories described in Sec. \ref{MGMI} (from  Ref. \cite{milgrom14b}), Eq. (\ref{jatare}) tells us that taking $V$ to be the mass-weighted, RMS velocity, and $m$ the total mass, we have
\beq  \eta\_V=\frac{4}{9}[1-\sum_p (m_p/M)^{3/2}]^2.   \eeqno{kaoper}
So $\eta\_V$ does depend on mass ratios in the system. But, in the common case where a system is made up of many light constituents -- such as a galaxy made of stars -- where $\sum_p (m_p/M)^{3/2}\ll 1$, $\eta\_V$ does take the universal value $\eta\_V=4/9$.

(d) In AQUAL and QUMOND, for all pure-disc (ideal) galaxies, if we take $\SB$ and $\SD$ as the central baryonic and dynamical surface densities, then $\eta\_\S=2$ is predicted \cite{milgrom16}.
\par
Central and important as these predicted strict relations are, we also seek definitions of such $\eta$ ratios that will form the basis for more general,  MOND primary prediction. Namely, definitions that result in predictions that are theory independent, and that
are applicable to (almost) any galaxy model, or, at least, to a large enough class, to afford meaningful comparison with observations. We can then relinquish strict universality, as long as the predicted values do not vary too much among theories and among system type. This would establish a tight correlation between the quantities in the numerators and denominators of $\eta$, as primary predictions of MOND that can be deduced from only the basic tenets\footnote{A meaningful prediction of such a correlations would result if the predicted scatter in $\eta$ is much smaller than the range of values of the correlated attributes that is accessible observationally.}.
\par
In what follows, I demonstrate the following with concrete examples:
(a) How, in a given theory, a strict (universal over all DML systems) relation that is predicted for a given choice of attributes in $\eta$ -- such as those in (a)-(d) above -- degrades into a correlation with scatter, when other attributes are used to define $\eta$.
(b) How some choices of the $\eta$ parameter lead to theory-dependent, and system-dependent predictions of its value, albeit with only a scatter ``of order unity''.
(c) How to define the system attributes that go into $\eta$ in such a way that a theory- and system-independent prediction of an order-unity scatter in $\eta$ follows from only the basic MOND tenets.
\subsection{Mass-velocity relations  \label{MV}}
The MASR, $\vinf^4=M\azg$ is the quintessential DML, mass-velocity relation, with $\eta\_V=1$ universally\footnote{It is the fact that in this case $\eta\_V$ is universal that constitute the quintessential prediction. The value $\eta\_V=1$ is imposed by our choice of normalization of the value of $\az$, or $\azg$. This value then enters all other MOND predictions.}.
It pertains to all isolated systems, whether they are characterized by low or by high accelerations within their bulk, and it involves only the asymptotic, circular speed of a test particle. There can be additional mass-velocity relation predicted by MOND (see below) that, despite similar appearance (which follows from the DML scaling properties), are very different, and quite independent of the MASR.
\par
Below, I discuss different aspects of such relations between the constituent masses of an isolated, self-gravitating object, in virial equilibrium, and some characteristic, intrinsic velocity. Unlike the MASR, the prediction of such a relation applies only to DML systems (which I assume in what follows), and employs, not the asymptotic, circular speed, but some bulk speed of the system's constituents.
\par
As I demonstrate in Sec. \ref{koka} below, it follows as a primary prediction -- i.e., from only the basic MOND tenets -- that with an {\it appropriate} choice of the velocity attribute in Eq. (\ref{conseq}), we have $\eta\_V=O(1)$ (namely, ``of order unity'') for all MOND theories, and all DML systems.
\par
But first I shall go through various predictions of such a relation, derived analytically, for different theories and different systems. These will give us a sense of what the exact value of $\eta\_V$ may depend on, and what may be meant by $\eta\_V=O(1)$.
\par
Start, as a benchmark, with the relation (\ref{kaoper}) between the total mass, $M$, and the mass weighted, RMS velocity dispersion.
 As described in Sec. \ref{MGMI}, $\eta\_V\equiv \s^4/M\azg=4/9$ is a prediction of a large class of MG theories,
when the system is made of a large number of small-mass constituents, so we can neglect $\sum_p (m_p/M)^{3/2}\ll 1$. \footnote{This holds, e.g., when the system is made of $N\gg 1$ masses $m\_p\sim M/N$, so this quantity is $\sim N^{-1/2}$.}
\par
But, even within such MG theories, the predicted value of $\eta\_V$ does depend on the choice of the velocity attribute. For example, as seen from (\ref{kaoper}) , a different value of $\eta\_V$ is predicted for a system with a small number of constituents. For a system of $N$ constituents of equal masses $M/N$, the value of $\eta\_V$ is reduced by a factor $(1-N^{-1/2})^2$. Or, if we have a system of many test particles, of
total mass $m$, around a central, large mass, $M\gg m$, and $\s$ is the RMS velocity of the test masses, then (\ref{kaoper}) gives $\eta\_V\approx 1$ (the MASR is a special case of this).
\par
\begin{figure}[ht]
	\centering
\includegraphics[width = 9cm] {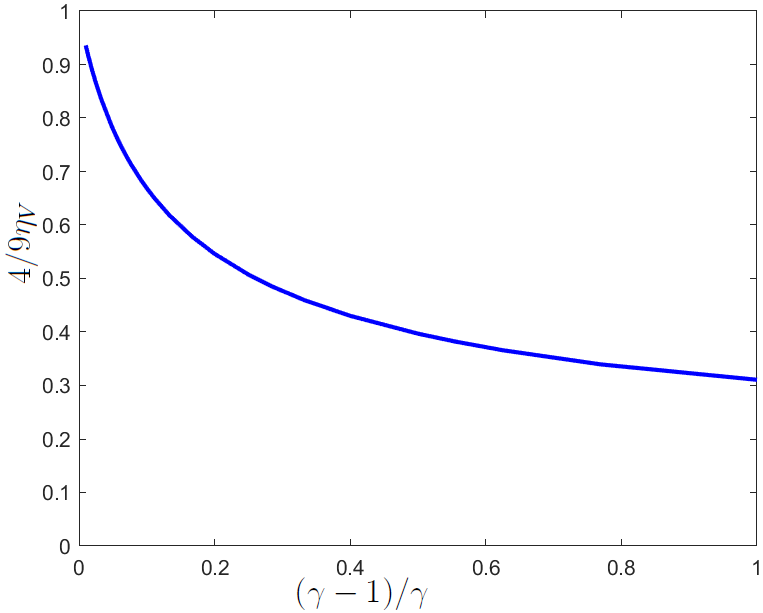}
\caption{From Ref. \cite{milgrom21}: Predicted correction --in MG theories -- to the benchmark value of $\eta\_V=4/9$, plotted vs the polytropic index $(\c-1)/\c$, for DML, isotropic polytropes, when we use $V= \sqrt{3}\s\_0$, where $\s\_0$ is the density-weighted, central line-of-sight velocity dispersion of the polytrope ($\sqrt{3}\s\_0$ is its 3-D equivalent).}
\label{Dn}
\end{figure}

Other departures from the benchmark value of $\eta\_V$ are predicted if we use other definitions of the characteristic $V$ in the definition of $\eta\_V$. Some examples are discussed in Ref. \cite{milgrom21}, where $\eta\_V$ is calculated for DML polytropes, for different definitions of $V$. As an example,
Figure \ref{Dn} shows the correction to the benchmark value, when we take $V$ in the definition of $\eta\_V$, as the central, line-of-sight, density-weighted velocity dispersion, $\s\_0$. For a meaningful comparison, we should use the three-dimensional equivalent of $\s\_0$, namely $\sqrt{3}\s\_0$.
We see that, even for the extreme case of an infinite polytropic index (a homogeneous sphere), the predicted $\eta\_V$ is about three times the benchmark value, becoming unity for $\c=1$ -- the isothermal-sphere case (for which $\sqrt{3}\s\_0$ equals the mass-weighted velocity dispersion).
\par
Additional insights can be gained by considering the predictions of the toy models \cite{milgrom25} described briefly in Sec. \ref{toy}, which are spanned by the parameter $1\le\b\le 2$.
One can derive the general virial relation in these models, for a DML, stationary system made of point masses $m\_p=\m\_p M$, where $M$ is the total mass
\beq  \av{\D r}^{4-2\b}\av{|\vr_p^{(2/\b)}|^{2\b}}=M\azg,   ~~~~  \av{\D r}\equiv \left(\sum_{q<p}\frac{\m_p\m_q}{|\vr_p-\vr_q|^{4-2\b}}\right)^{-1/(4-2\b)}.  \eeqno{gatarew}
Here, $\av{\D r}$ is some measure of the mass-weighted-average inter-constituent distance -- a measure of the system size -- and $\av{|\vr_p^{(2/\b)|}}$ is the mass-weighted average of this fractional, time derivative of the trajectory -- a nonlocal quantity. The left-hand side is the 4th power of some characteristic velocity, $\s$.
With these theory-dependent choices of $m=M$, and $V=\s$, in the definition (\ref{conseq}) of $\eta\_V$, we have $\eta\_V=1$.
\par
{\it While these definitions lead to system-independent values of $\eta\_V=O(1)$, they do not seem useful, as they involve quantities that are practically impossible to measure, and which, to boot, are strongly theory dependent.}

\par
For $\b=1$ (the linear model), relation (\ref{gatarew}) gives
\beq  \eta\_V\equiv \frac{(\D r)^2\av{\va^2}}{M\azg}=1,   \eeqno{madret}
where
\beq \av{\va^2}=M^{-1}\sum_p m_p \va_p^2,~~~~~~~~\D r\equiv \left[\sum_{q< p}\frac{\m_p \m_q}{|\vr_p-\vr_q|^2} \right]^{-1/2}.  \eeqno{mindas}
\par
For $\b=2$ we have the simple result
\beq \eta\_V\equiv \av{v^4}/M\azg=\frac{1}{2} (1-\sum_p\m_p^2),   \eeqno{relgeq}
where $\av{v^4}$ is the mass-weighted average of $|\vv|^4$.

\subsubsection{Binary relative velocity -- circular orbits}
The relation between the masses and the relative velocity in an isolated, DML binary (the separation does not enter because of the SI of the DML) is another example of a mass-velocity relation. The velocity parameter could be some long-time average of the relative velocity over the trajectory.
Generally, we expect the resulting $\eta\_V$ to depend not only on the mass ratio, but also on dimensionless attributes of the trajectory (e.g., the maximum-to-minimum-separation ratio). But, for the case of circular orbits, we can derive some analytic results in different theories, as follows: Take $V=V\_{12}$, the relative velocity, and $M$ the combined mass; so,
\beq  \eta\_V\equiv V_{12}^4/M\azg.   \eeqno{matdus}
The large class of MG theories described in Sec. \ref{MGMI} predict, from Eq. (\ref{kaoper}):
\beq \eta\_V= \frac{4}{9}\left(\frac{1-\m_1^{3/2}-\m_2^{3/2}}{\m_1\m_2}\right)^2,   \eeqno{manish}
where $\m_i\equiv m_i/M$. $\eta\_V$ varies between $\approx 0.6$ and $1$.
\par
The class of MI models described in Ref. \cite{milgrom23} predict
\beq \eta\_V= (\m_1^{1/2}+\m_2^{1/2})^2,   \eeqno{shumda}
which varies between $1$ and $2$.
\par
The $\b$ models predict
\beq \eta\_V=[\m_1^{1/(2\b-1)}+\m_2^{1/(2\b-1)}]^{2\b-1}.  \eeqno{kacha}
(For all cases, we get for $\m\_2\ll\m\_1$, $\eta\_V=1$, which is the MASR.)

\subsubsection{A ``universal'' DML $M-\s$ correlation  \label{koka}}
I demonstrated above the exact constancy of the ratio $\eta\_V\equiv V^4/m\azg$, for any given MOND theory, within any family of homologous, virialized, DML mass distributions, and for any definition of $m$ and $V$ that are subject to the scalings (\ref{scasca}) or (\ref{scabla}).
\par
I also demonstrated above that given a theory, one may be able to define attributes $m$ and $V$ that predict a system-independent value of $\eta\_V$.
What we want, however, is a definition of $m$ and $V$ that are reasonably amenable to observational determination, and -- also important -- that yield a predicted DML correlation between $m$ and $V$ that is independent of the MOND theory, so that we can test MOND generally without committing to specific theories.
\par
To derive such correlation, we need to make the following concessions: (a) we cannot take any arbitrary mass and velocity attributes to correlates, and (b) we have to be satisfied with a strong correlation, instead of an exact relation between the mass and the velocity attributes.
\par
We shall see that to get such a correlation we cannot rely only on the SI of the DML: we shall also need to confine ourselves to MOND theories that have a Newtonian limit at high accelerations, and in which the MOND-to-Newtonian transition occurs across an acceleration range $\sim\az$. In other words, we shall need all the MOND tenets (i)-(iv) listed in Sec. \ref{axioms}. The $\b$ models discussed in Sec. \ref{toy} do not satisfy axioms (iii) and (iv).
\par
Consider any family of virialized, homologous, DML systems.
Define system attributes $M$, $\s$, and $R$ such that: (a) in the Newtonian regime, these quantities satisfy the Newtonian virial relation $\s^2\approx MG/R$, and (b) the MOND-to-Newtonian transition occurs when $R\approx\rM\equiv (MG/\az)^{1/2}$.
Then, in the definition (\ref{conseq}) of $\eta\_V$, take $m=M$, and $V=\s$.
\par
In a given MOND theory, all the DML members of the family share the same value of $\eta\_V=\eta\_V\^M$. While $\eta\_V\^M$ can depend on the theory and on the family, we want to show that it must be $O(1)$ for the above choice of attributes.
\par
The family is spanned by the two parameters $\k$ and $\l$ in Eqs. (\ref{scasca}-\ref{scabla}), which control the mass and the size of the systems. The MOND-Newtonian borderline is some line $\k\propto \l^2$ in the $(\k,\l)$ plane, which corresponds to $R=\rM(M)$. Moving in this plane towards this borderline, axiom (ix) tells us that for a system in the family at the borderline, both the MOND and the Newtonian relations hold to order unity. This means that on one hand, $\eta\_V$ still has approximately the value $\eta\_V\^M$, and, on the other hand, that the Newtonian virial relation also holds to order unity for $R=\rM$. Putting these two together, we have that $\eta\_V\^M=O(1)$, as required.
\par
To recapitulate,
{\it the Newtonian limit plays an important role in constraining DML dynamics, independently of the specific MOND theory at hand}. This is because (a) Newtonian dynamics is a single, unique theory that all MOND theories must flow to in the limit of high accelerations, (b) because it is required to be near enough the borderline on the Newtonian side, and (c) because the DML scalings relate DML dynamics near the borderline to the whole DML region.

\subsubsection{Notes on comparison with observations}
We saw above, in Eqs. (\ref{kaoper}) (\ref{gatarew}) (\ref{madret}) (\ref{relgeq}), examples where specific theories predict universal, constant values of $\eta\_V$. These all involve the total mass, but they involve different definitions of the velocity attribute. These velocity attributes are, however, practically impossible to determine observationally, and in testing the MOND predictions, as well as the Newtonian ones, one has to employ proxies, for example, velocity dispersions that are (a) the line-of-sight component, instead of the three-dimensional one, (b)  values derived from just a small sample of the system's constituents (e.g., using the globular clusters around a galaxy, and not stars), and (c) values that do not weigh properly the different contributions (e.g., luminosity-weighted, instead of mass-weighted dispersions). These are only some of the systematics that stand in the way of direct comparison of the predictions with the observations.
\par
I mention this in the present context to indicate that we do not degrade the comparison very much by having predictions of only a correlation with
$\eta\_V=O(1)$, since the scatter introduced in the observed correlation due to the various systematics and measurement errors can be larger than that introduced by possible variations in the predicted $\eta\_V$ among systems.
\par
Indeed, the fortunate fact is that such correlations can be tested over ranges of values of the numerator ($V^4$) and the denominator ($m$), which define $\eta\_V$, that are much larger -- many orders larger -- than the scatter introduced by both the systematics and the leeway in the predicted values of $\eta\_V$.
\par
\begin{figure}[!ht]
\begin{tabular}{lll}
\includegraphics[width=0.7\columnwidth]{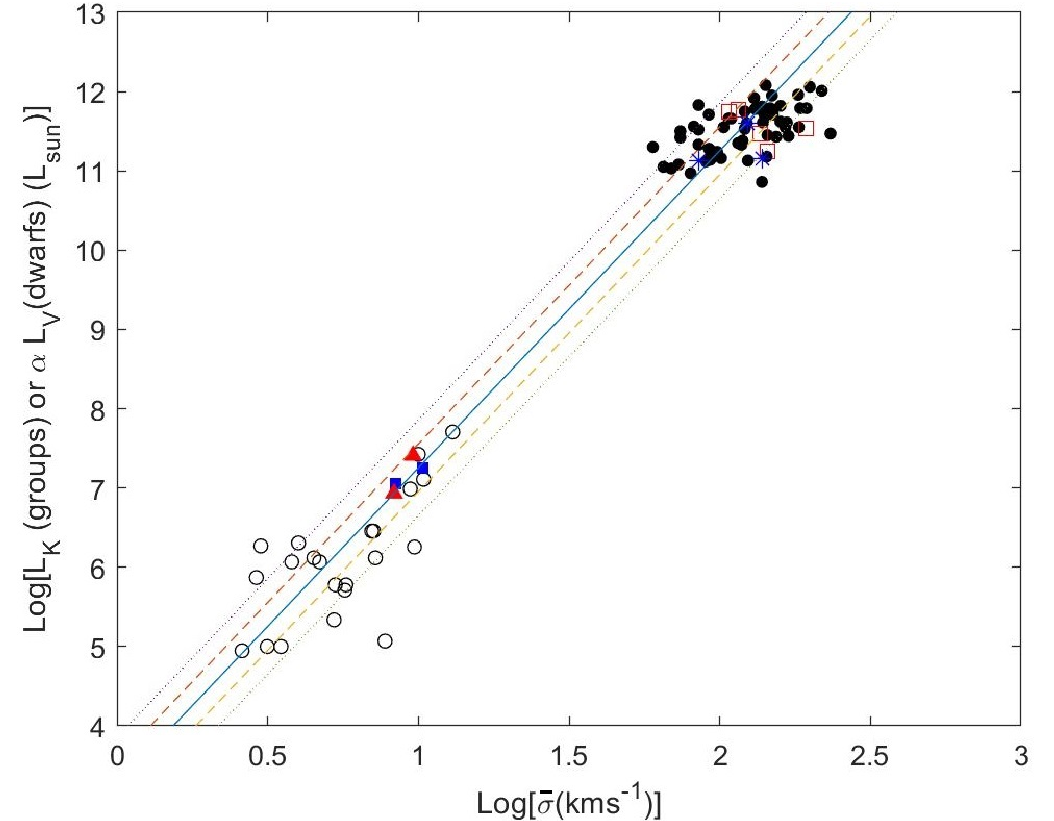}
\end{tabular}
\caption{From Ref. \cite{milgrom19}: $L\_K/L\_{K,\odot}$ for the groups, and $\a L\_V/L\_{V,\odot}$ for dwarfs ($\a=2/0.7$), plotted vs. the line-of-sight velocity dispersion.  The solid line is the prediction for $\eta\_V=4/9$ [see the paragraph containing Eq. (\ref{kaoper})] and taking $\s=\sqrt{3}\s\_\|$ ($\s\_\|$ being the literature value of the line-of-sight, global velocity dispersion)), for $M/L\_K=0.7s.u$ for the groups, and for the more appropriate V-band $M/L\_V=2s.u.$ for the dwarfs. The dashed lines are for 0.5 (upper) and 2 (lower) times these $M/L$ values, and the dotted lines are for 0.25 (upper) and 4 (lower) times these values. Details on the sources of the data, and further details are given in Ref. \cite{milgrom19} (caption of Fig. 6 there).
\label{msigmagroups}}
\end{figure}

As an example, Fig. \ref{msigmagroups} shows a plot, from Ref. \cite{milgrom19}, of the luminosity (a proxy for the mass) vs a line-of-sight velocity dispersion in dwarf spheroidal galaxies, and in galaxy groups -- two classes of DML systems\footnote{While the groups have typical internal accelerations of $(0.5-3)\times 10^{-2}\az$ (see Fig. 7 if Ref. \cite{milgrom19}), the dwarfs,  typically, have accelerations  $(2-15)\times 10^{-2}\az$ (see Table 1 of Ref. \cite{mcgaugh13}).}, with, otherwise, very different attributes.
\par
Details are given in the caption and in Ref. \cite{milgrom19}. I only show the Figure here to demonstrate that
while the sources of scatter and systematics for the dwarfs and for the groups are quite different, they are still much smaller than the span of the observed correlation, resulting in a very meaningful test despite all the scatter.

\subsection{Correlation between baryonic and dynamical surface densities  \label{CSDR}}
Now consider the parameter, $\eta\_\S$, which encapsulates -- and generalizes -- the DML branch of the MOND prediction of a tight correlation between baryonic and dynamical surface densities. As in the case of $\eta\_V$, within any family of homologous, isolated, DML mass distributions\footnote{We can now drop the requirement that the systems be virialized, because the dynamical mass distribution can be determined from the kinematics of test particles -- bound and unbound -- that need not be part of the system.}, $\eta\_\S$ is predicted to be strictly constant for any given definition of the baryonic and the dynamical surface densities, that are subject to the scalings (\ref{scasca}) or (\ref{scabla})\footnote{This means that they are expressed in terms of the scalable quantities. For example, we can use the mean baryonic surface density within half the projected half-mass radius.}.
\par
However, it is impracticable (at present, at least) to restrict such a test to an homologous family of galaxies. And, we expect that with general definitions like this, $\eta\_\S$ takes very different values for different families. And, as with $\eta\_V$, we expect such general definitions of $\eta\_\S$ to be strongly dependent on theory.
\par
So, again, we have to make some concessions in order to obtain a testable universal correlation, both in restricting the definition of the surface densities that enter $\eta\_\S$, and in accepting that we end up with a correlation (i.e., with a predicted scatter in $\eta\_\S$), not a strict relation.
\par
An extensive discussion of such a primary prediction has been given in Ref. \cite{milgrom24}, with references to other works on the subject. Here I give a succinct account.
\par
As in the previous section, I start with some concrete example to serve as a benchmark, against which the possible variety in $\eta\_\S$ can be assessed. It was shown in Ref. \cite{milgrom16} that the MG theories AQUAL and QUMOND predict the following relation: For all pure-disc, DML mass distributions (e.g., pure-disc galaxies), modelled as an axisymmetric thin mass distribution, one has $\eta\_\S\equiv \SDz/(\SBz\SM)^{1/2}=2$, where
$\SBz$ is central surface density of the baryonic disc, and $\SDz$ is the dynamical, central surface density. The latter is defined as the dynamical column density along the symmetry axis of the disc. In Ref. \cite{milgrom24}, I showed that this result does not hold exactly even in more general MG theories, e.g., those within the class discussed above, in Sec. \ref{MGMI}.
\par
In MI theories, the concept of dynamical densities, and surface densities, is not even well defined generally; so, neither is $\eta\_\S$. The reason is that the concept of dynamical density requires an acceleration field $\vg(\vr)$ to be defined, from which the dynamical density, $\r\_D\equiv -\div\vg/4\pi G$, is gotten. But in MI theories, the test particle acceleration depends not only on position \cite{milgrom23}; so an acceleration field is not defined. It is, however, possible to define such quantities, and construct a meaningful definition, with well defined predictions, if we confine ourself to certain derivation of dynamical quantities; for example, if we restrict definitions of such quantities to those derived from rotation curves, and use strictly the accelerations derived for circular orbits.
\par
As with $\eta\_V$, the predicted value of $\eta\_\S$ depends not only on the underlying theory, but within a given theory it depends on the specific mass distribution.
For example, within AQUAL and QUMOND, the above benchmark value for discs is not predicted for spherical systems, for which it takes up a higher values, which, furthermore, depend on the exact radial density distribution (see details in Ref. \cite{milgrom24}).
\par
\begin{figure}[ht]
	\centering
\includegraphics[width = 11cm] {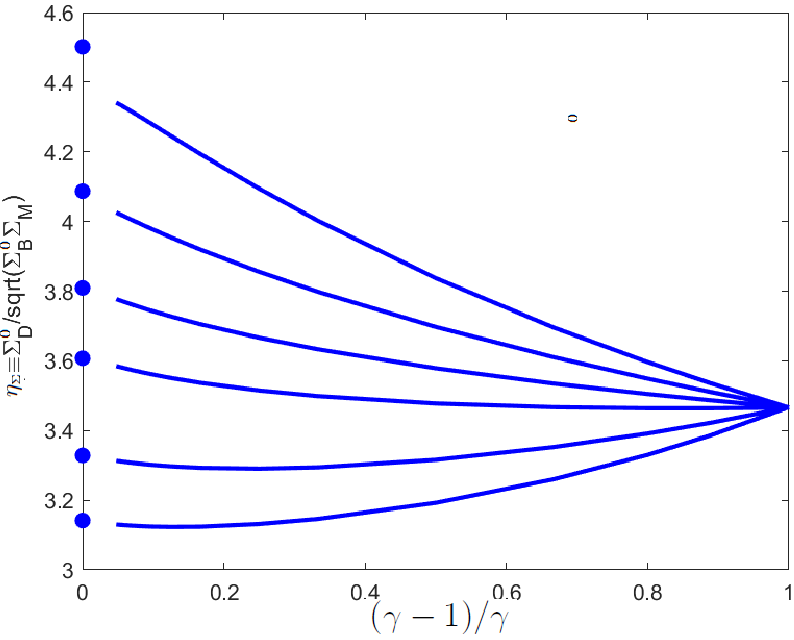}
\caption{$\eta\_\S$,  predicted in AQUAL and QUMOND, constructed from the baryonic and dynamical central surface densities for anisotropic, DML polytropes [vs. $(\c-1)/\c$, $\c$ is the polytropic index], for anisotropy ratios (from bottom to top) $\b=-0.5,~-0.3,~-0.1,~0,~0.1,~0.2$. Dots: isothermal spheres. $\c\rar\infty$ corresponds to homogeneous spheres for all $\b$ values.}
\label{yag}
\end{figure}

Figure \ref{yag} (from Ref. \cite{milgrom21}) exemplifies this for anisotropic, DML polytropes, as predicted by AQUAL/QUMOND. We see that for the full range of polytropic index, and for a large range of velocity-anisotropy ratios, we have $3\le\eta\_\S\le 4.5$, to be compared with the benchmark value, $\eta\_\S=2$, predicted for discs.
\par
As mentioned above, in MI theories we cannot define an acceleration field from which to define a dynamical density for general baryonic mass distributions. However, for spherical systems, we can define a specific (spherical) dynamical mass distribution as that which, in Newtonian dynamics, gives rise to the observed, or predicted, rotation curve. With this definition, the predicted dynamical density, in pure MI theories, is identical with that predicted by AQUAL/QUMOND, since, from the general theorem in \cite{milgrom94}, the predicted rotation curves are the same in all these theories, and given by Eq. (\ref{munater}).

\subsubsection{Universal DML $\SD-\SB$ correlation  \label{kiper}}
To obtain a system-, and theory-independent correlation, we proceed in a similar manner to what we did in Sec. \ref{koka}, based on the same assumptions (see Ref. \cite{milgrom24} for details). Again, we start with a homologous family of mass distributions. We do not choose the baryonic and dynamical surface-density attributes arbitrarily -- as we can do when considering an single homologous family -- but take them such that (a) the dynamical surface density reduces to the baryonic one for Newtonian systems, and (b) The MOND-to-Newtonian transition {\it for these attributes} is defined to order unity in the $(\k,\l)$ plane. This means that $\SD$ has to become approximately $\SB$ when the whole systems transitions between the limits. This means, in turn, that they both have to be of some global significance in the systems, such as the global average, or the central value if this is a good indicator of where we are on the MOND-Newtonian scale\footnote{An example of an extreme case where the argument does not work, is the when the surface densities are defined by the column densities along a line outside the baryonic mass. In this case $\SB=0$, but $\SD$ is finite in MOND, and $\eta\_\S$ becomes infinite.}.
With these requirements satisfied, as above, the $\eta\_\S$ value for the whole family of DML distribution is of the same order as the Newtonian value of the ratio, which is unity.

\end{document}